\documentstyle[11pt,aaspp4]{article}  

\begin{document}   

\title{References for Galaxy Clusters Database}
\author{M.\ Kalinkov, I.\ Valtchanov, I.\ Kuneva}
\affil{Institute of Astronomy, Bulgarian Academy of Sciences, 72
  Tsarigradsko Chaussee blvd, BG-1784 Sofia, Bulgaria}
\authoremail{markal@astro.bas.bg}
%
%


\begin{abstract}          
  A bibliographic database will be constructed with the purpose to be
  a general tool for searching references for galaxy clusters. The
  structure of the database will be completely different from the
  available now databases as NED, SIMBAD, LEDA.  Search based on
  hierarchical keyword system will be performed through web interfaces
  from numerous bibliographic sources -- journal articles, preprints,
  unpublished results and papers, theses, scientific reports. Data
  from the very beginning of the extragalactic research will be
  included as well.
  
  References for galaxy clusters (RGC) is continuation of a previous
  project for collecting all published information for galaxy
  clusters. There are neither previous attempts nor projects for
  compiling such database for galaxy clusters and our effort will be
  to include into the database all of the available bibliography
  through our system of keywords until the end of 1999. Now over
  3\,000 entries are included into the preliminary version of the
  database.
\end{abstract}

\section{Introduction}
Clusters of galaxies are the largest gravitationally bound systems in
the Universe. They are the most effective tracers of the large scale
structure of the Universe, as well as of the formation history and
evolution of the largest mass scales.

About 16\,000 clusters of galaxies are known at present and there is a
large amount of observational data for about 2--3\,000.

It is true that some information on clusters (catalogs, lists, etc.)
can be retrieved from any astronomical data center, but it is
impossible more specific information to be obtained. The only
exception is NASA/IPAC Extragalactic Database (NED). But despite of
the tens of thousands references stored in NED, there is no easy way
to find where, e.g. photometry is published for a given cluster.

The main aim of our project is to give the opportunity for queries
where data for any observation or theoretical consideration is
published or stored. It is reasonable to test the basic concepts which
will be incorporated in the database using A/ACO clusters of galaxies
(Abell 1958, Abell et al. 1989). And this work concerns mainly the
data for these clusters.

\section{Catalogs of clusters of galaxies. Identifications}

The first catalog of clusters of galaxies was published by Abell
(1958) for $\delta \geq -27^{\circ}$ and extended by Abell, Corwin \&
Olowin (1989) for $\delta \leq -17^{\circ}$. Thus both catalogs are
unique in their coverage of the entire sky. We denote the clusters
from these catalogs with prefix ``A'' (which is not the same as in
NED). There are A1 $\div$ A4076 clusters, about 4050 different.
Moreover, ACO contains about 1150 supplementary (S) southern clusters
which are either not rich enough or at a great distance to be included
in the main catalog.

Many clusters are found by Zwicky et al. (1961--68) but for $\delta
\geq 0^{\circ}$. Their total number is 9133 (denoted with prefix
``ZC'').

There are other lists of groups and clusters:

\begin{deluxetable}{lcc}
\label{table1}
\tablehead{
\colhead{Reference} & \colhead{Prefix} & \colhead{Number of Objects}}
\startdata
\multicolumn{3}{c}{Visual procedure}\nl
Klemola (1969) & Kl & 44\nl
Sersic (1974)  & Ser& several tens\nl
Rose (1976)    & Ro & 124\nl
Duus \& Newell (1977) & DN & 710\nl
Snow (1978)    & Sn & 29\nl
Braid \& MacGillivray (1978) & BMc & $\sim 400$\nl
Quintana \& White (1990) & QW & 267\nl
\multicolumn{3}{c}{Semi-automatic algorithm}\nl
Shectman (1985) & Sh & 646\nl
\multicolumn{3}{c}{Abell-like automatic algorithm}\nl
Lumsden et al. (1992) & ED & 737\nl
Dalton et al. (1997) & APM & 957\nl
\enddata
\end{deluxetable}

Various visual searching procedures for clusters result in a
subjective catalogs.  While the catalog of Shectman is the first one
where a ``blind''computer search was performed among the famous Lick
counts of galaxies (Shane \& Wirtanen 1967, Seldner et al. 1977).

The Abell-like algorithm catalogs are entirely automated and so they
are entirely objective.
 
Clusters of galaxies are found also by Jackson (1982), Gunn, Hoessel \&
Oke (1986), Willick (1991) and Postman et al. (1996).

Some designations of clusters are proposed and used before A/ACO
catalog publication and some of them have clear correspondence to the
objects in the A/ACO catalog. For example, A3266 = Ser040/6 =
DN0431-616 = APM510 and references for all these objects must point to
A3266. So it would be impossible to retrieve the information available
for a given A/ACO cluster without making first a cross-identifications
of clusters among all available catalogs or lists. Some
identifications are presented in the original ACO catalog with some
misprints or errors.

We present here a new cross-identifications (Appendix 1) between A/ACO
clusters and the other clusters. Abell--Zwicky cross-ident\-ifi\-ca\-tions
are taken from Kalinkov et al. (1997, 1998). Appendix 2 contains some
essential notes about A-ZC identifications.

\section{Two ways to retrieve references}

The basic idea is to assign two lists to each reference -- the first
contains the A/ACO clusters and the second contains the keywords
describing observational data or results.

Two ways for references retrieval for any cluster are possible --
using a glossary of keywords or using a thesaurus which is created for
that particular need of clusters of galaxies.

The present version of this project operates with a glossary, given in
Appendix 3 (where names and designations as Coma, Perseus, Virgo,...
 Shapley Scl, Ind Scl, Psc-Cet void, Hydra A, \dots 3C28, 3C40, NGC 1272, NGC
1275 \dots are not included).

Using a thesaurus of the kind compiled by Shobbrook \& Shobbrook
(1993) would be convenient but it is very hard to create a new
thesaurus for such a narrow field like clusters of galaxies. If a
usual thesaurus is available then it would be quite simple to find any
reference assigned with a narrower term. But it is not possible to use
a narrower terms for clusters of galaxies, when there are hierarchical
structures (double galaxies, groups, superclusters, substructures and
so on). Two examples are given in Appendix 4 -- for primary terms (PT)
CLUSTERS OF GALAXIES and SUPERCLUSTERS OF GALAXIES. Evidently, the
broader term (BT) is STRUCTURES. The corresponding narrower terms (NT)
as well as related terms (RT) are presented. The main difficulty is
that many NT denoted with ``$\ast$'' -- Catalogs, Crossing time,
Searching radius, \dots, have meaning which is different for different
PT. So the $\ast$ figures as a comment or a scope note (SN).  As a
conclusion, a general retrieval of references is possible for
combinations of NT and PT.

\section{Current status}

\begin{description}
\item[(i)] 3000 references are included -- journal articles, workshops,
  symposia, colloquia papers, thesis works.
\item[(ii)] References for all A/ACO clusters of galaxies, together
  with their cross-identification (Appendix 1) may be retrieved at
  present. There are about 5200 clusters.
\item [(iii)] The glossary is given in Appendix 3.
\end{description}

\section{Future developments}

\begin{description}
\item[(i)] Actualizing the references. At the end of 1999 we expect to
  entry about 12\,000 titles.
\item[(ii)] Including non A/ACO clusters -- about 16\,000 objects. 
\item [(iii)] Extending the glossary.
\item [(iv)] Creation of thesaurus oriented to the field of clusters
  of galaxies.
\item [(v)] Free web access to the database
\end{description}

\section*{Acknowledgements}

This project was initiated ten years ago as a part of a project {\it
  ``Reference Catalog of Clusters of Galaxies''}. The creation of the
database for clusters of galaxies is a complex and time consuming
work, having in mind all the recent developments and the pouring
observations of clusters of galaxies. Many colleagues have helped us
and it is impossible to mention all of them. But we would like to
thank particularly H.~Andernach, H.~Corwin, G.~Dalton, G.~de
Vaucouleurs, J.~Huchra, W.~Keel, R.~Nichol, F.~Ochsenbein, A.~Omont,
G.~Paturel, M.~Postman, M.~Ramella, H.~van der Laan, R.~Williams.

This research has made use of the NASA/IPAC Extragalactic Data\-ba\-se
(NED) which is operated by the Jet Propulsion Laboratory, California
Institute of Technology, under contract with the National Aeronautics
and Space Administration.

This work was supported by the National Research Fund of the Bulgarian
Ministry of Education, Science and Technology (contract F721/1997).

\clearpage
\baselineskip 2pt

\begin{verbatim}
         A P P E N D I X  1
         ------------------
                                       
A/ACO    z    nz  Cross-identifications
=====+======+====+======================
A0001 0.1249   1  ZC0022 
A0003 0.1012   1  ZC0029 
A0005             ZC0037 
A0006             ZC0039 
A0007 0.1073   1  ZC0048 
A0009 0.0949   1  ZC0052 
A0013 0.0943  37  APM021 Sh500 
A0014 0.0657   8  APM029 ED418 Sn01-05 
A0015 0.1211   2  APM030 ED419 
A0016 0.0838   1  Sh502 ZC0071 
A0017 0.0882   1  ZC0069 
A0021 0.0946  11  ZC0104 
A0022 0.0646   5  APM050 ED437 
A0023 0.1052   1  Sh508 ZC0105* 
A0024 0.1332   1  ZC0106* 
A0025             ZC0108 
A0026 0.1462   2  ZC0109 
A0027 0.0540   2  APM062 Sh511 
A0028 0.1845   2  ZC0121 
A0029             ZC0131 
...
S0002 0.0647   1  ED393 BM433 
S0003             ED395 
S0004             DN0001-702 
S0008             DN0004-681 
S0012 0.0513   3  APM012 ED407 BM279 
S0017             APM022 ED414 BM282 DN0011-380 
S0019             DN0013-681 
S0020             DN0013-700 
S0021 0.0383   2  QW001 
S0027             QW002 DN0016-697 
S0028 0.0535   1  Sh506 
S0034             BM052 
S0035 0.1232   5  DN0020-394 
S0037 0.0528  18  DN0021-425 
S0041 0.0498   1  ED443 
S0043             APM066 
S0045 0.0248   1  DN0022-571 Ro025 
...

 * - comment in Appendix 2 for A-ZC correspondence

\end{verbatim}

\clearpage
\begin{verbatim}
         A P P E N D I X  2
         ------------------
                                       
A/ACO Comments
=====+==========================
A0023 hfZC0112
A0024 fZC0075
A0055 hfZC0173
A0064 +A0065+A0067=ZC0194?
A0065 +A0064+A0067=ZC0194?
A0067 +A0064+A0065=ZC0194?
A0071 heA0077=heZC0213,hbZC0181
A0075 heZC0217
A0077 heA0071=eZC0205
A0095 +A0101=ZC0254
A0101 +A0095=ZC0254
A0102 hbZC0265
A0132 fZC0322
A0142 fZC0368
...
...

   hb - half background
   hf - half foreground
   he - half equidistant
    f - foreground

\end{verbatim}

\clearpage

\begin{verbatim}
         A P P E N D I X  3
         ------------------

                GLOSSARY

1st,2nd... - First,second...rank g  gamma rays
21 cm                               gas
absorption lines                    GINGA
AGN - active galactic nuclei        globular cls
alignment                           gravitational lensing
angular momentum                    H0
arc                                 HI
arclet                              HII
associations                        HALPHA
background photometry               HR - Hertzsprung-Russell diagram
bar                                 HST - Hubble Space Telescope
BCG - bright cluster g              Hubble diagram
binary model                        I(s) - irregular(s) g(s)
black holes                         ICG - intracluster gas
blue arcs                           ICM - intracluster medium
blue gs                             identification
blue objects                        IGG - intergalactic gas
BM - Bautz-Morgan classification    IGM - intergalactic medium
BO - Butcher-Oemler effect          image
CCD                                 infall
CCD photometry                      IR - infrared
catalog                             IRAS
cD,cD2                              isophote photometry
cD LF - LF of cD gs                 isopleth map
cl(s) - cluster(s)                  IUE
cl detection                        jet
classification                      KARA - Karachentsev
CO - CO observations                kinematics
colors                              Lick - Lick observatory counts
compact gs                          luminosity
complex                             LF - luminosity function
cooling flow                        luminosity segregation
core                                LALPHA
correlation                         Malmquist bias
correlation function                map
COSMOS                              mass
counts                              mass/energy
crossing time                       M/L
cusp                                mass segregation
D25 - 25 mag. diameter              membership
db - dumb-bell                      model
density                             morphology
disk gs                             m-z - magnitude-redshift relation
distance                            NELGs - narrow emission line gs
DM - dark matter                    [OIII]
double gs                           O - optical
dust                                O identification
dynamics                            O position
dwarfs                              orbits
dEs - dwarf ellipticals             PA - positional angle
E(s) - elliptical(s)                peculiar
E&S0                                photometry
E+cD                                polarization
elongation                          polar-ring gs
emission line gs                    pole
emission lines                      profile
equidensities                       PUMA observations
EXOSAT                              QSO
evolution                           R - radio
Faraday rotation                    radius
[FeX]                               ring gs
FIR - far infrared radiation        RLF
formation                           ROSAT
g(s) - galaxy(-ies)                 rotation
g content                           S(s) - spiral(s)
g mass                              Scl(s) - supercluster(s)
Scott effect                        Tifft effect
segregation                         TF - Tully-Fisher relation
shell gs                            UBV
shape                               UV
simulations                         velocity
size                                velocity dispersion
SN - supernova                      velocity distribution
spectra                             velocity flow
spectroscopy                        velocity segregation
star formation                      VLA
statistics                          void
stellar population                  wavelet
Stroemgren                          wedge
structure                           X - X-ray radiation
subclustering                       X core
substructure                        XLF
superposition                       XRLF - X-ray and radio LF
Sy - Seyfert                        X models
SZ - Sunyaev-Zeldovich effect       X Scl
tail                                X subclustering
tidal effect                        z - redshift

\end{verbatim}

\clearpage

\begin{verbatim}
         A P P E N D I X  4a
         -------------------

PRIMARY TERM: CLUSTERS OF GALAXIES

BROADER TERM: STRUCTURES

                NARROWER TERMS:

 1st, 2nd, ..., 10th galaxy     *Luminosity
 Abell clusters                 *Luminosity function
*Alignments                     *Mass
 APM clusters                   *M/L
 BM classification              *Membership
*cD galaxies                    *Morphology
*Catalogs                        Orbits of galaxies
*Content                        *Radio data
*Correlation functions          *Relaxation time
*Counts                         *Richness
*Crossing time                  *Rotation
 D galaxies                     *Searching procedures
 Density cusps                  *Shape
*Distances                      *Simulations
*Dynamics                       *Size
 ED clusters                     Subclusters
*Evolution                      *Substructures
*Formation                      *Types
*Gas                            *Velocity dispersion
 Intracluster matter            *Velocity distribution
*Kinematics                     *X-ray data
 Lick clusters                   Zwicky clusters

                RELATED TERMS:
                
        Galaxies
        Great Attractor 
        Mergers
        Missing mass
        Peculiar velocities
        Scott effect
        Selections
        Sunyaev-Zeldovich effect
        Superclusters
        Velocity field
        Virial theorem

\end{verbatim}
\clearpage

\begin{verbatim}
         A P P E N D I X  4b
         -------------------

PRIMARY TERM: SUPERCLUSTERS OF GALAXIES

BROADER TERM: STRUCTURES

                NARROWER TERMS:

        *Catalogs
        *Crossing time
        *Correlation functions
        *Counts
        *Dynamics
        *Evolution
        *Formation
         Great Attractor
        *Kinematics
         Local Supercluster
        *M/L
        *Maps
        *Mass
        *Membership
        *Multiplicity
        *Radio data
        *Searching procedures
        *Shape
        *X-ray data

                RELATED TERMS:

        Clusters
        Selections
        Virial theorem

\end{verbatim}

\end{document}